\documentclass[letterpaper,12pt]{article}

\pdfoutput=1

\usepackage[margin=1in,letterpaper]{geometry}
\usepackage{booktabs}
\usepackage{courier}
\usepackage{graphicx}
\usepackage{listings}
\usepackage{program}
\usepackage{seqsplit}
\usepackage{afterpage}
\usepackage{algorithm}
\usepackage{textcomp}
\usepackage{longtable}
\usepackage{enumitem}

\begin{document}

\title{freebase-triples: A Methodology for Processing the Freebase Data Dumps}

\author{Niel Chah \\
University of Toronto}
\date{December 2017}

\maketitle

\begin{abstract}
The Freebase knowledge base was a significant Semantic Web and linked data technology during its years of operations since 2007. Following its acquisition by Google in 2010 and its shutdown in 2016, Freebase data is contained in a data dump of billions of RDF triples. In this research, an exploration of the Freebase data dumps will show best practices in understanding and using the Freebase data and also present a general methodology for parsing the linked data. The analysis is done with limited computing resources and the use of open-source Unix-like tools. The results showcase the efficiency of the technique and highlight redundancies in the data, with the possibility of trimming nearly 60\% of the original data. Once processed, Freebase's semantic structured data has applications in other prominent fields, such as information retrieval (IR) and knowledge-based question answering (KBQA). Freebase can also serve as a gateway to other structured datasets, such as DBpedia, Wikidata, and YAGO. 
\end{abstract}

\section{Introduction}

In this research, we explore a relatively short-lived knowledge base, Freebase, that made up a significant part of the Semantic Web and linked data field. The term \textit{Semantic Web} was proposed by Tim Berners-Lee et al. in 2001 to describe a system of structuring information on the Web to be intelligible for machines \cite{berners2001semantic}. The term \textit{linked data} was also coined by Berners-Lee in 2006 to emphasize the potential for data from one source to link to other datasets in a systematic manner \cite{bernersLD}. During a similar time frame, recent years have seen the proliferation of technologies that advance the state-of-the-art in large-scale analysis of massive data sets. Among others, technologies such as Hadoop \cite{White2012} and Spark \cite{Shanahan2015} have facilitated such analyses. At the same time, they often require advanced technical knowledge and computing resources. 

This paper's main research contribution is a framework to parse and explore the defunct Freebase data dumps with limited computing resources. First, we provide a historical overview of Freebase from its launch, acquisition by Google, and its eventual shutdown. We also present a brief survey of current uses of the data in such fields as information retrieval (IR) and knowledge-based question answering (KBQA). Next, with particular attention to its unique schema, we present a methodology designed to effectively parse the billions of RDF triples in the Freebase data dumps. Finally, we end with a discussion of our findings and consider limitations of the Freebase archives.

\section{Background}

\subsection{Freebase: Inception, Acquisition, and Shutdown}

\begin{figure}[h]
	\centering
	\caption{A screenshot of freebase.com on May 2, 2016 before it was shut down.}
	\includegraphics[width=0.75\columnwidth]{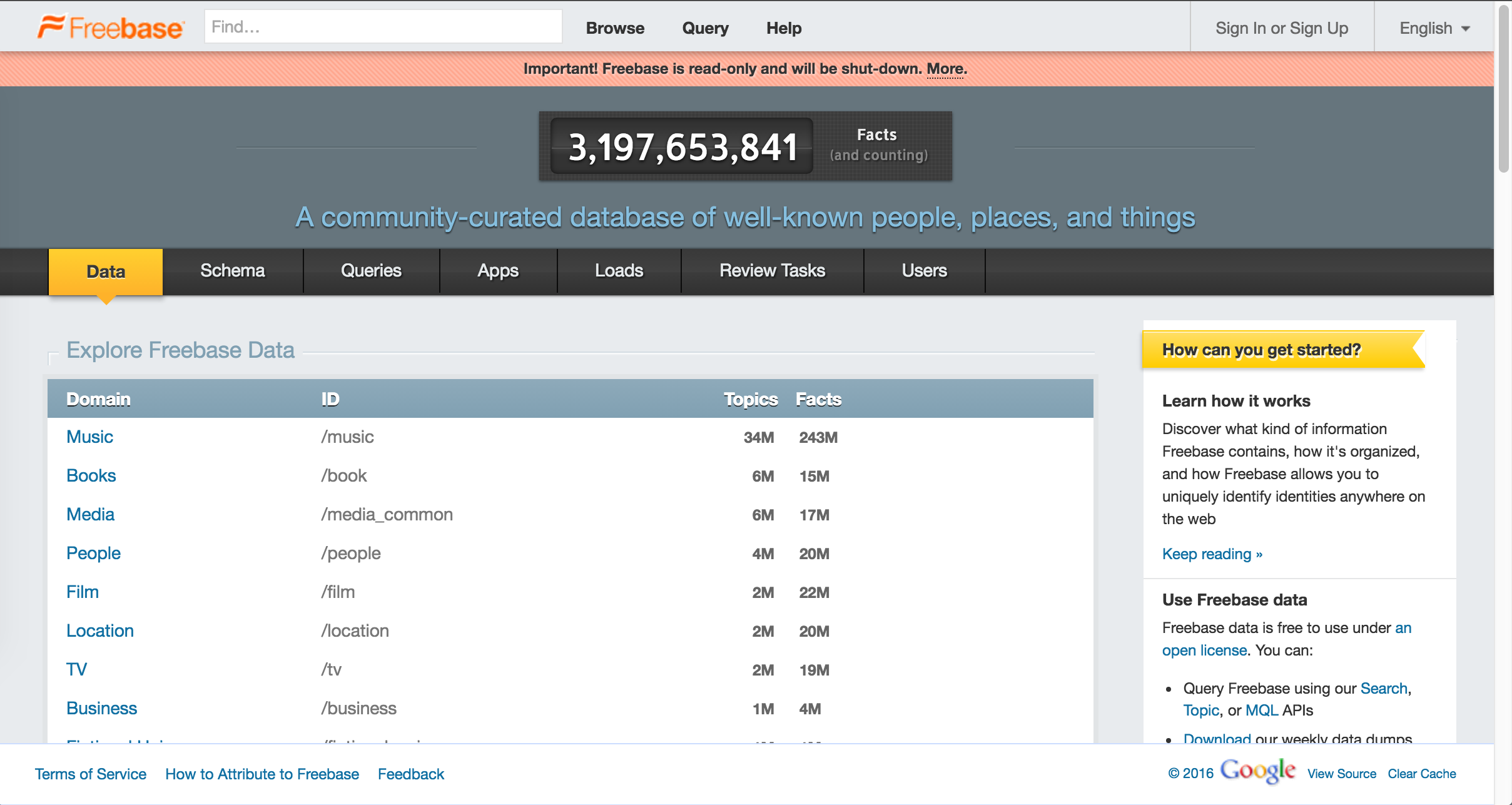}
	\label{fig1}
\end{figure}

In 2007, Freebase (see Figure \ref{fig1}) was launched by Metaweb as an open and collaborative knowledge base \cite{Bollacker2008}. On the website freebase.com, users could register a free account and edit the data of various entries, creating linkages between entities in the data. This was similar to the kind of editing and data entry that was possible on Wikipedia. A significant difference was that while Wikipedia predominantly consists of free-form text suitable for an encyclopedic article with links to other resources, Freebase was exclusively concerned with encoding links and relationships between entities in the knowledge base.

In 2010, Metaweb was acquired by Google for an undisclosed sum that gave it possession of the Freebase knowledge base and its then 12 million entities \cite{Menzel2010}. Freebase was used at Google to power parts of their internal Google Knowledge Graph, which supported Google Search features such as Knowledge Cards or Panels \cite{PellissierTanon2016}. On December 16, 2014, it was announced on the Freebase Google Plus community that Freebase would be gradually shut down over the next six months, and its data would be migrated to the Wikidata platform \cite{Gplus2017}. In reality, the Freebase service remained open for the community and was finally shut down on May 2, 2016 \cite{Douglas2016}. From this date onward, all freebase.com URLs began to redirect to the Google Developers page for the once active knowledge base.

\subsection{Freebase in Current Usage}

Freebase data is applied in a variety of experiments and research in the fields of information retrieval (IR), knowledge-base question answering (KBQA) and even artificial intelligence (AI). Recent research applies advances in neural networks to Freebase's rich semantic data. For instance, Bordes and colleagues propose a neural network architecture to embed information from knowledge bases such as Freebase into a vector space in order to facilitate information retrieval \cite{bordes2011learning}. In their research, the authors emphasize the hidden potential in applying semantic data to AI fields such as natural language processing and computer vision. In another application of neural networks, Dong and colleagues implement a multi-column convolutional neural network using Freebase data to answer natural language questions \cite{dong2015question}. Further work in question answering using Freebase data was conducted by Yao and Van Durme, Bordes et al., and Fader et al. \cite{yao2014information,bordes2014question,fader2014open}. Knowledge graph data has also been conceived as tensor decompositions in order to derive structured rules and apply data mining techniques as done by Narang, and Papalexakis et al. \cite{Narang2016,papalexakis2016tensors}.

There is a large body of research on data profiling to gain insights \cite{abedjan2015profiling,naumann2014data}. Researchers have profiled and analyzed the Freebase knowledge base in the past when freebase.com was still active. However, there has been relatively less research on Freebase data itself after the website was shutdown and preserved as a data dump. Färber et al. conducted a comprehensive survey of knowledge bases, including Freebase, by comparing them along a number of criteria such as the domains they cover and their schema \cite{Farber2015}. In further research, Färber et al. analyze the same knowledge bases along an array of data quality dimensions, such as their suitability for particular use cases \cite{Farber2016}. Another notable study of Freebase data was done by Pellissier Tanon et al. in their detailed evaluation of the migration of Freebase data to Wikidata and the resulting challenges \cite{PellissierTanon2016}. Bast et al. also explore Freebase through a unique methodology \cite{Bast2014}. Although the precise methodology is different from the approach inthis paper, the use of Unix tools to explore RDF data was also done by Bendiken \cite{Bendiken2010}. This paper's methodology builds on these existing works and provides new contributions through a framework to understand the archive of Freebase data.

\section{Freebase}

\subsection{Freebase Schema}

The specific terminology and schema used by the Freebase knowledge base will be briefly explained. A thorough understanding of these aspects will contribute to a more effective workflow in processing the data dumps. First, the most prominent terms will be introduced through the following example. In Freebase, a distinct entity or object, such as the notable film director Steven Spielberg, is called a \textit{topic}. Each topic is associated with a unique machineId identifier (often abbreviated as \textit{mid}) that is in the format ``\texttt{/m/ + \{alphanumeric\}}'', such as \texttt{/m/abc123}. The topic for Steven Spielberg is said to be a member of a class by saying it ``has certain \textit{type(s)}''. In this example, the famous film director will have the Person type (written in a human-readable format as \texttt{/people/person}) and the Film Director type (\texttt{/film/film\_director}). Under each type, further granular data is represented through \textit{properties}, such as his Place of Birth (\texttt{/people/person/ place\_of\_birth}) and the films he directed (\texttt{/film/film\_director/films}). A property can link a topic to a value (\texttt{/people/person/date\_of\_birth} - \texttt{Dec 18, 1946}) or other topics (\texttt{/people/person/place\_of\_birth} points to the mid for ``Cincinnati, Ohio''). 

\begin{figure}[h]
	\centering
	\caption{A simple graph of triples}
	\includegraphics[width=0.5\columnwidth]{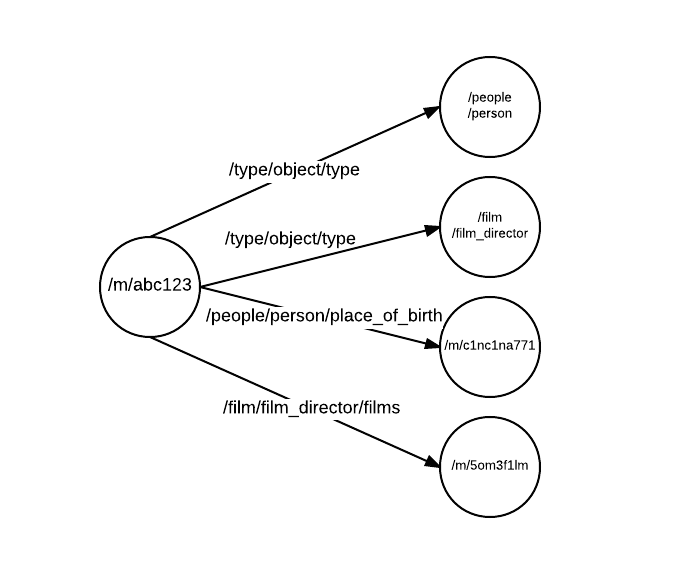}
	\label{fig2}
\end{figure}

The Freebase schema (the system specifying the types and properties) and data (expressing the facts about topics) are expressed in a triples format. This data can be visually represented as a directed graph (see Fig. \ref{fig2}). The Resource Description Framework (RDF) is a widely used W3C specification to encode triples data \cite{RDF2014}. A triple links a (1) \textit{subject} through a (2) \textit{predicate} to a (3) \textit{object}. This format intuitively mirrors how a single triple expresses a fact as a natural language sentence: Johnny Appleseed (\textit{subject}) likes (\textit{predicate}) apples (\textit{object}). The Spielberg example could thus be written with the following triples:

\begin{lstlisting}
# /type/object/type indicates an entity's types
/m/abc123, /type/object/type, /people/person
/m/abc123, /type/object/type, /film/film_director

# These triples express facts about /m/abc123 
/m/abc123, /people/person/place_of_birth, /m/c1nc1na771
/m/abc123, /film/film_director/films, /m/5om3f1lm
\end{lstlisting}

The schema of Freebase types and properties, often called an \textit{ontology} in the Semantic Web context, has an interesting human-readable aspect that borrows from the Unix-like use of forward slashes ``/'' in directories. In the Freebase ontology, there are a number of domains that are centered on specific subject matters. For example, there is a domain for People and another domain for Films, expressed as the \texttt{/people} and \texttt{/film} domains respectively. Within each domain, there are type(s) for the classes of entities within a domain. Thus, there is a Person type, expressed by a human-readable ID as \texttt{/people/person}, and a Deceased Person type, expressed as the \texttt{/people/deceased\_person} type. In the film domain, notable types include the Film Director \texttt{/film/film\_director} and Film \texttt{/film/film} among others. Within each type in turn, there are properties to capture more granular data after one further ``/'', such as the Date of Birth \texttt{/people/person/date\_of\_birth}.

\section{Methodology}

\subsection{Freebase Data Dumps Characteristics}

The Freebase data dumps are available for download in a N-Triples RDF format.\footnote{freebase.com} According to the Google Developers website, the Freebase Triples version of the data dumps is composed of 1.9 billion triples in a 22 GB gzip file (and over 200 GB uncompressed) \cite{FBdatadumps}.

To accommodate the data into the N-Triples RDF format, a number of changes were made by Google that differentiate it from the  original Freebase encoding that was used on freebase.com. First, all machineId MIDs and schema that are usually in the format \texttt{/m/abc123} or \texttt{/people/person} are instead written as \texttt{/m.abc123} and \texttt{/people.person} respectively, so that any forward slashes after the first instance are converted to full stops. All objects in the data are also encoded with a full URL path, such as \seqsplit{http://rdf.freebase.com/ns/m.abc123}, so that it is globally compatible with other RDF linked data datasets. Although this format is necessary to conform to the RDF standard, certain simplifications will be applied for the purposes of this project.

With regards to the overall format, the data file has a consistent structure. Each row (or each triple) of data is terminated by a full stop and newline ``. \textbackslash n''. Each element of the triple is enclosed in angle brackets ``\textlangle \textrangle'' and delimited by a tab character ``\textbackslash t''. A single line of RDF triples from the data dumps is shown as follows (with the different parts of the triple separated for readability). In the following implementation section, a number of pre-processing steps will be applied to transform the dataset into a more workable form.

\begin{lstlisting}
<http://rdf.freebase.com/ns/g.112ygbz6_>
<http://rdf.freebase.com/ns/type.object.type>
<http://rdf.freebase.com/ns/film.film> .
\end{lstlisting}

The values that are linked to through properties are also in the form of data types, such as strings, dates, and numeric values. Strings are encoded with the value enclosed in double quotations and appended with the ``@'' character and the ISO 639-1 standard language code: ``string''@en \cite{LOClangs}. Numeric values are encoded as is, without additional markup. Date values are appended with additional markup text in the form of ``\seqsplit{\textasciicircum\textasciicircum http://www.w3.org/2001/XMLSchema\#date}''.

\subsection{Implementation}

The data dumps were parsed using a variety of shell scripts and command line tools run on a MacBook Pro (Early 2015 model, 2.7 GHz Intel Core i5, 8 GB RAM) and an external hard drive (Seagate 1 TB). One of the contributions of this paper is the use of tools available on Unix-like systems, such as \texttt{awk, cat, cut, gawk, grep, gsed, less, more, parallel, pv, sed, sort, wc, zless, zmore}, and \texttt{zgrep}. These command line tools were chosen for their performance and wide availability on operating systems, such as Linux distributions and macOS. The code is available in its entirety on GitHub \cite{nchah2017fbtriples}.\footnote{https://github.com/nchah/freebase-triples}.

\begin{figure}[h]
	\centering
	\caption{Overview of the stages in the code}
	\includegraphics[width=.5\columnwidth]{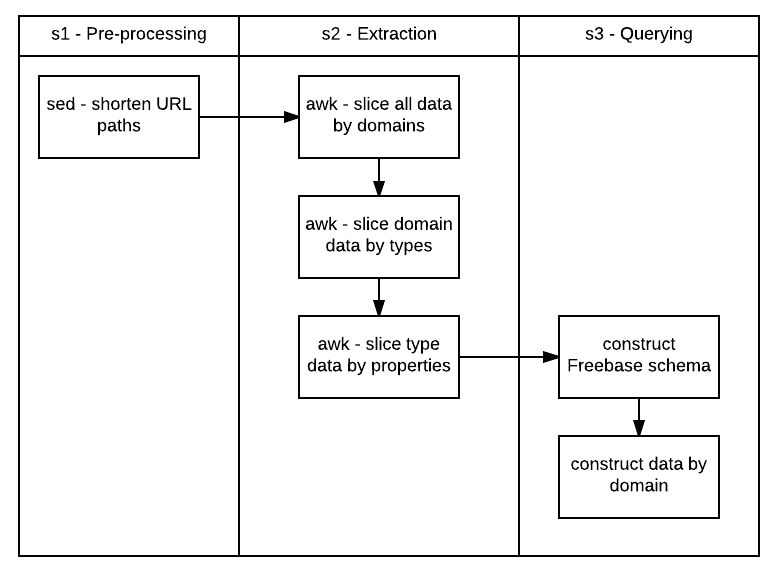}
	\label{fig3}
\end{figure}

The codebase is conceptually divided into three stages that process the data dumps: (1) pre-processing, (2) extraction, and (3) query stages (see Fig \ref{fig3}). In the first pre-processing stage, the sed utility shortens the full URL path by removing the protocol, subdomain, and domain so that only the path remains. That is, ``http://rdf.freebase.com/ns/m.abc123'' becomes \texttt{/m.abc123}. The latter variant is significantly more human-readable. Running this script on the data dumps reduced the storage size from approximately 425.2 GB to 219 GB while still preserving the triples count. Further optional scripts are included in the code to remove the angle brackets and convert full stops to forward slashes.

The scripts in the second extraction stage used \texttt{awk, cat, cut, grep, parallel}, and \texttt{sort} tools to slice out portions of the data dumps based on the Freebase schema and RDF structure \cite{tange2011gnu}. A \textit{slice} is defined as a subset of the RDF triples where the triple's predicate (middle) term is part of a unique domain, type, or property. This is an intuitive slicing method since the predicate term is the edge or link between the subject and object nodes if triples are conceptualized as a graph. Thus, a slice contains all the triples for each unique kind of edge in the graph. The slicing is fastest by first obtaining the unique predicate terms and then iteratively slicing out the domains in order of decreasing triples count so that the most frequent predicates are sliced out first. This slicing can continue so that increasingly narrower domain, type, and property slices can be created (see Figure \ref{fig4}).

Another priority was the extraction of topic ``identifier'' slices. For this paper, these ``identifier'' triples are defined as the triples that express a Freebase mid's names, aliases, descriptions, types, and keys (which are links to external authority control or databases such as IMDb or Wikipedia). The extraction of these triples is an integral step in determining a Freebase topic's coherent ``identity'' as a representation of a real world entity. Additional scripts that followed this methodology extracted the globally unique predicates, types, and the underlying schema.

In the third query stage, a workflow was established for exploring a specific domain. The code focuses on combining the triples from the slices created in the previous stage in order to reconstruct the topics in a domain. This methodology facilitates deeper analysis of the Freebase data by organizing the unwieldy data dumps into smaller human understandable slices. With the data dumps processed in this way, it should be possible to achieve the following objectives.

\begin{itemize}[noitemsep]
\item identify redundancies in the current data and propose improvements,
\item obtain comprehensive statistics on the data
\item guide future research that intends to use Freebase semantic data.
\end{itemize}

\afterpage{
\begin{figure*}[p]
	\centering
	\caption{Overview of the slicing workflow}
	\includegraphics[width=\textwidth,height=.2\textheight,keepaspectratio]{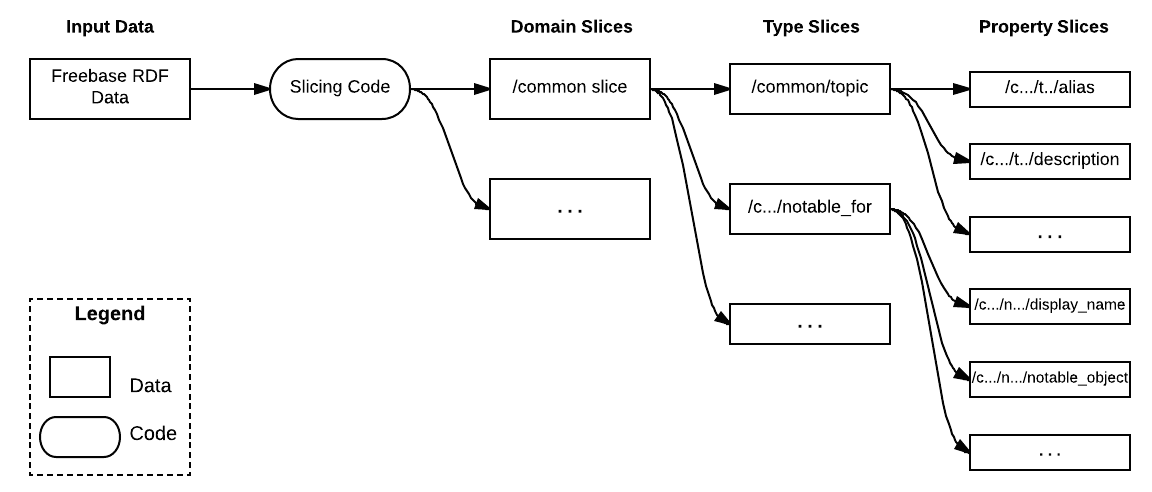}
	\label{fig4}
\end{figure*}

\begin{figure*}[p]
	\centering
	\caption{Distribution of slices by number of triples}
	\includegraphics[width=\linewidth,height=.3\textheight,keepaspectratio]{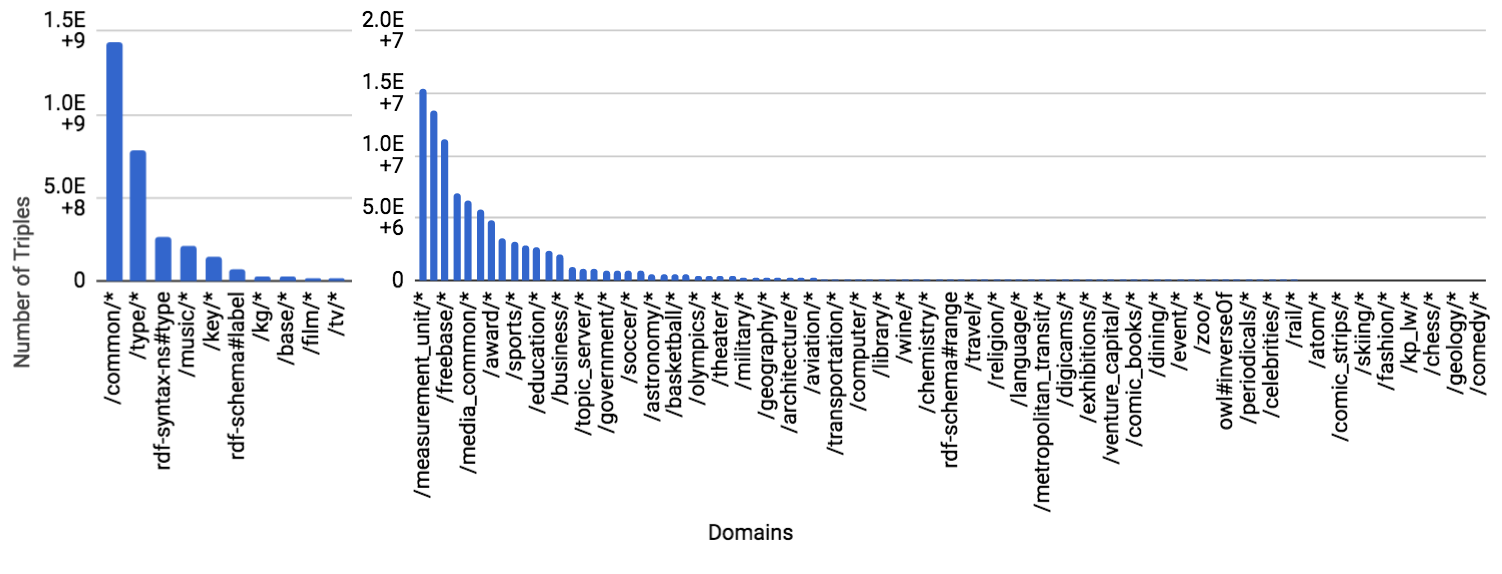}
	\label{fig5}
\end{figure*}

\begin{figure*}[p]
	\centering
	\caption{Distribution of slices by processing runtimes}
	\includegraphics[width=\linewidth,height=.3\textheight,keepaspectratio]{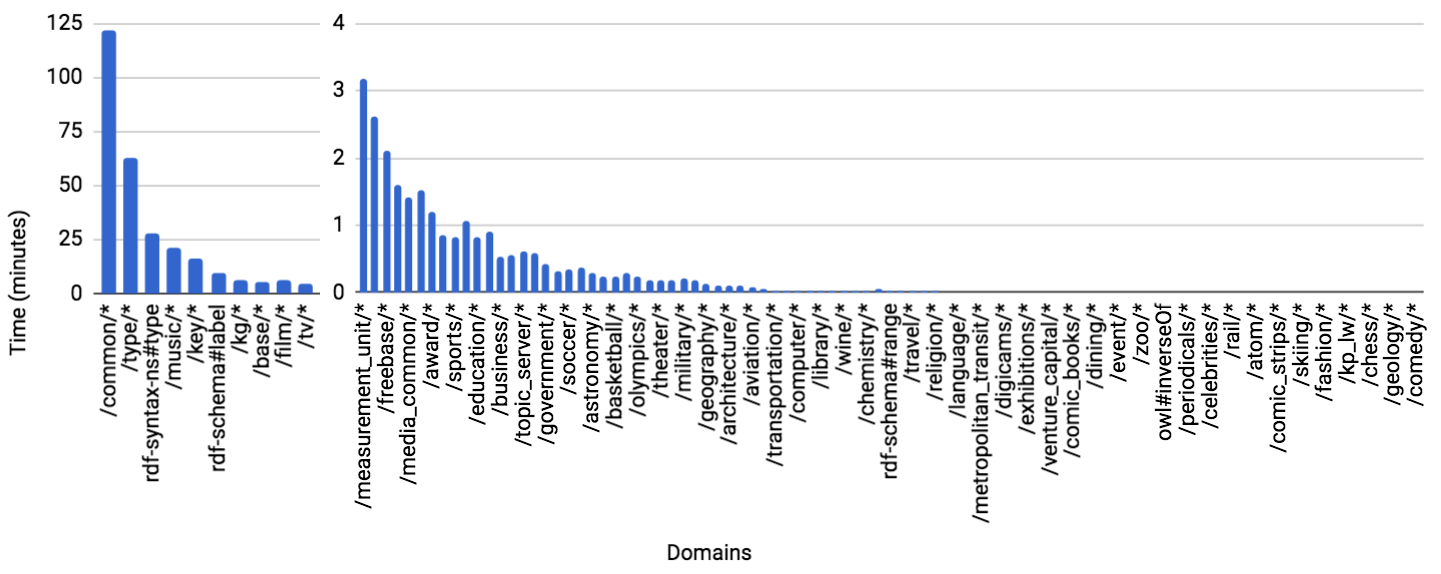}
	\label{fig6}
\end{figure*}
\clearpage
}

\section{Findings}

\subsection{Evaluating Slices: Redundancy and Performance}

By slicing out the triples belonging to the largest domains first, the subsequent slicing operations were run on increasingly smaller slices of RDF triples (see Fig. \ref{fig5} and Fig. \ref{fig6}). With this methodology and the computing resources of a personal laptop, over 95\% of the triples were sorted by slicing the first 10 domains in 282 minutes, or under 5 hours (see Table \ref{table1}). With the bulkiest slices extracted first, the each remaining slice took under 5 minutes to process, with increasingly faster times as the slicing progressed. In total, 315 minutes were spent to slice all domains.

\begin{table}[h]
	\centering
	\caption{A selection of the Freebase slices by processing runtime}
	\label{table1}
\begin{tabular}{ |l|l|r|r|r| } 
\hline
Name & Domain (or Predicate) & Number of Triples & \% & Slicing Runtime (min) \\
\hline
common & /common/* & 1,429,443,085 & 45.658 & 121.8262 \\
\hline
type & /type/* & 788,652,672 & 25.191 & 63.1563 \\
\hline
owl-type & rdf-syntax-ns\#type & 266,321,867 & 8.507 & 28.0785 \\
\hline
music & /music/* & 209,244,812 & 6.684 & 20.8632 \\
\hline
key & /key/* & 149,564,822 & 4.777 & 15.8327 \\
\hline
owl-label & rdf-schema\#label & 72,698,733 & 2.322 & 9.7133 \\
\hline
kg & /kg/* & 30,689,453 & 0.980 & 6.6167 \\
\hline
base & /base/* & 24,063,303 & 0.769 & 5.5669 \\
\hline
film & /film/* & 17,319,142 & 0.553 & 5.9002 \\
\hline
tv & /tv/* & 16,375,388 & 0.523 & 4.6004 \\
\hline
Totals & & & 95.963 & 282.1544 \\
\hline
\end{tabular}
\end{table}

Obtaining more granular slices for types and properties produces interesting findings that point to possible optimizations that reduce redundancies in the original data. For instance, five slices are based on predicates involving the ``http://www.w3.org'' Web Ontology Language (OWL). These slices contain triples where the predicate is in the form ``\seqsplit{\textlangle https://www.w3.org/2000/01/rdf-schema\#label\textrangle}'' which is equivalent to the Freebase \texttt{/type/object/name} property. There are exactly 72,698,733 triples with this OWL predicate that duplicates the same number of Freebase counterparts. This repetition is also found with the URL for the ``\seqsplit{\textlangle https://www.w3.org/2000/01/rdf-syntax-ns\#type \textrangle}'' predicate which mirrors Freebase's \texttt{/type/object/type}. This kind of redundancy accounts for 10.83\% of the triples in the data dumps. 

An aspect of the Freebase system that may be considered an additional redundancy is the encoding of triples with predicates in a \textit{forward} and \textit{reverse} direction. A triple such as ``\texttt{/m/abc123, /type/object/type, /people/person}'' expresses an entity's types in a \textit{forward} direction. The same semantic information with a predicate in the \textit{reverse} direction would express a type's instances as ``\texttt{/people/person, /type/type/instance, /m/abc123}''. These redundancies with the predicate \texttt{/type/type/instance} make up 8.5\% (266,257,391 triples) of the data dumps. In total, nearly one fifth (10.83 + 8.5 = 19.33\%) of the data can be trimmed from the data dumps.

Interestingly, the Common (\texttt{/common/*}) and Type (\texttt{/type/*}) domains together make up over 70\% of the entire data dumps. Within this subset, 40.91\% of triples (1,280,720,680 triples) involve the \texttt{/common/notable\_for/*} and \texttt{/common/topic/notable\_for} predicates. The schema descriptions for these properties cannot be found in the data dumps, but it can be inferred that the properties express what an entity is ``notable for''. For example, Barack Obama would be notable as a U.S. President. However, this data is expressed verbosely using a mediator object:

\begin{lstlisting}
1: /m/02mjmr (Barack Obama), /common/topic/notable_for, /g/125920_xt(mediator)
2: /g/125920_xt, /common/notable_for/display_name, "US President"@en # "en" and other languages...
3: /g/125920_xt, /common/notable_for/object, /government/us_president
4: /g/125920_xt, /common/notable_for/predicate, /type/object/type
5: /g/125920_xt, /common/notable_for/notable_object, /government/us_president
\end{lstlisting}

Although the existing implementation may have made sense for the Freebase infrastructure at the time, it is a verbose representation in the data dumps. A more efficient expression of this notability relationship could be made by linking the subject directly to the \texttt{/type/type} object, such as \texttt{/government/us\_president}. This change would only consume 0.98\% (30,696,375 triples) of the current data. In total, the removal of redundant data (19.33\%) and the change to the notability data (40.91\%) would result in the removal of nearly 60\% of the original data dumps (59.26\% = 19.33\% + (40.91\% - 0.98\%)).

\subsection{Data Analytics: Domain and Identifier Slices}

The ``identifier'' triples make up 16.31\% of all triples (see Table \ref{table2}), with the \textit{type} and \textit{key} triples making up the lion's share. From these statistics, it is possible to estimate the total number of entities represented in the data dumps. Under a loose definition, a Freebase mid must have at least a \textit{name} in one language namespace in order to represent a real world entity. Using an \texttt{awk} script and the \texttt{wc} utility, the total number of unique MIDs in the subject position of the name slice (that is, where the predicate is \texttt{/type/object/name}) was obtained. This resulted in an estimated count of 51,847,135 possible topics.

Applying this workflow to obtain statistics on domains is also useful. As explained in the schema portion of this paper, all types and properties are part of a top-level domain. The identifier triples are part of the Type (/type) domain or the Common (/common) domain. In addition to these domains, Freebase has a number of domains on diverse subject matters ranging from American Football to Zoos. A total of 105 domains were found (see Table \ref{table3}, and the Appendix for the full table).

\begin{table}[h]
	\centering
	\caption{Freebase ``identifier'' slices}
	\label{table2}
\begin{tabular}{ |c|l|r|r| } 
\hline
Slice & Predicate & Number of Triples & \% of All \\
\hline
All data & & 3,130,753,066 & 100 \\
\hline
name & /type/object/name & 72,699,101 & 2.32 \\
\hline
type & /type/object/type & 266,321,867 & 8.51 \\
\hline
keys & /type/object/key & 146,583,100 & 4.68 \\
\hline
desc & /common/topic/description & 20,472,070 & 0.65 \\
\hline
akas & /common/topic/alias & 4,611,150 & 0.15 \\
\hline
Total & &510,687,288 & 16.31 \\
\hline
\end{tabular}
\end{table}

\begin{table}[h]
	\centering
	\caption{A selection of Freebase slices, by subject matter domains}
	\label{table3}
\begin{tabular}{ |l|l|r|r| } 
\hline
Name & Domain (or Predicate) & Number of Triples & \% of All \\
\hline
american\_football & /american\_football/* & 278,179 & 0.009 \\
\hline
amusement\_parks & /amusement\_parks/* & 22,880 & 0.001 \\
\hline
architecture & /architecture/* & 253,718 & 0.008 \\
\hline
astronomy & /astronomy/* & 556,381 & 0.018 \\
\hline
automotive & /automotive/* & 46,543 & 0.001 \\
\hline
\end{tabular}
\end{table}

\subsection{Reconstructing Freebase Schema: The Bicycles Domain}

The schema (or \textit{ontology}) of a domain is also expressed in RDF triples, and can be reconstructed from the data dumps. In order to explore this reconstruction in further detail, it will be helpful to explore the Bicycles (\texttt{/bicycles}) domain due to its relatively small size (at 22 KB and 313 triples). An exploration of this domain will proceed without any existing knowledge of the domain or its schema to demonstrate how to make sense of a domain from triples alone.

First, statistics on the general shape of the domain can be obtained by finding the number of unique objects in each of the subject, predicate, and object positions of the triples. Next, the count of unique predicates is supplemented with an output of the actual unique predicates. These predicates are comprised of the properties (keeping in mind the \texttt{/domain/type/property} structure of the schema) for a domain. The types can be inferred quite easily by making a ``hop'' backwards by one forward slash (or full stops). In the Bicycles domain, the following types and properties were found:

\begin{lstlisting}
</bicycles.bicycle_model.manufacturer>
</bicycles.bicycle_model.speeds>
</bicycles.bicycle_model.bicycle_type>
</bicycles.bicycle_manufacturer.bicycle_models>
</bicycles.bicycle_type.bicycle_models_of_this_type>
\end{lstlisting}

After the most important components of the schema are established, further searches were done to compile details on the types and properties. Each type and property is also linked to a mid, name, description, and further schema specific parameters. The mid acts as a unique identifier, in addition to the human-readable ID (i.e. the form written in \texttt{/domain/type/property}). A sample of the output is displayed below for the Bicycle Type type (e.g. mountain bikes, tandem bikes, etc.).

\begin{lstlisting}
# /bicycles/.../bicycle_type has mid /m.05kdnfz 
</bicycles.bicycle_model.bicycle_type>, 
	</type.object.name>, 
		"Bicycle type"@en 
</m.05kdnfz>, 
	</common.topic.description>, 
		"The type or category of bike, eg. mountain bike, recumbent, hybrid"@en
\end{lstlisting}

A full overview of the Bicycle domain reveals three main types: Bicycle Model, Bicycle Type, and Bicycle Manufacturer. According to the schema details, the Bicycle Model is connected to the Bicycle Type via the \texttt{/bicycles/bicycle\_model/bicycle\_type} property. For instance, the mid for a bicycle model, \texttt{/m/s0meB1k3} will have the type \texttt{/bicycles/bicycle\_model} and it will be linked through the property \texttt{/bicycles/ bicycle\_model/bicycle\_type} to the mid \texttt{/m/m0unta1nB1k3}. The triples that express this relationship are shown below.

\begin{lstlisting}
# /m/s0meB1k3 is a "Bicycle Model"
/m/s0meB1k3, /type/object/type, /bicycles/bicycle_model
/m/s0meB1k3, /bicycles/bicycle_model/bicycle_type, /m/m0unta1nB1k3

# /m/m0unta1nB1k3 is a "Bicycle Type" 
/m/m0unta1nB1k3, /type/object/type, /bicycles/bicycle_type
/m/m0unta1nB1k3, /bicycles/bicycle_type/bicycle_models_of_this_type, /m/s0meB1k3
\end{lstlisting}

\section{Discussion}

\subsection{Efficient Data Mining Guided by Schema}

The pre-processing, extraction, and query stages in the code were guided by an understanding of Freebase's unique characteristics. Awareness of implementation details such as the format of the triples, the unique notation and encoding of Freebase, and the structure of the ontology contributed to code that could process the data dumps most efficiently.

Furthermore, the conceptualization of different categories of predicates (``identifier'', ``domain'', etc.) resulted in a system of organized slices. The identifier triples were shown to be integral to reconstructing the topics, facts, and even schema in Freebase. The domain-based triples led to an understanding of the knowledge base as consisting of different subject matters. Data on a specific domain, such as Bicycles, could be examined for its schema and its facts. From there, more nuanced discussions on the coverage of the data (does it cover all kinds of bicycles?), accuracy of the data (is the data correct?), the decisions in the schema (does it make sense to capture only the Bicycle Type and Manufacturer?), and other questions can be considered.

\subsection{Data Mining with Limited Resources (on a Budget)}

It was also found that the current implementation using shell programming languages and tools is suitable for data mining projects in environments with limited computing resources. For example, with slight modifications to the code and considering the optimizations of the redundancies found in this research, the current framework could even be run (slowly) on an affordable Raspberry Pi. An advantage of utilizing these predominantly open-source technologies is the greater accessibility of data mining technology to people without access to extensive resources.

This challenge could have been avoided entirely by utilizing greater computing resources either in the form of higher performance computers or cloud services, such as Amazon Web Services or Google Cloud. However, high performance computing resources are not an efficient replacement for a framework that processes data with an understanding of the underlying schema. With the cloud option, it should be noted that there may be concerns with dependency on a specific cloud service. While switching to another cloud service is possible, the costs associated with billing, maintenance, and debugging on the new platform should also be considered.

\subsection{Limitations of the Data Dumps}

It is also important to consider how the Freebase data dumps leave out crucial information that would otherwise be available on the operational knowledge base website. Although the Freebase triples are contained in a RDF data file, the freebase.com community functioned similarly to the active community of editors and reviewers on Wikipedia. Much like Wikipedia, freebase.com was accessible through a web UI where users could contribute semantically linked data through manual edits or programmatic processes. In this way, each user's editing of a single triple had a wealth of associated metadata such as the provenance (the user or process responsible for the data), the timestamp (when the data was added), and whether the edit was adding or deleting data (potentially a signal for edit vandalism). This metadata is an important part of the overall Freebase system that is missing from the data dump and not easily recoverable.

In addition to the lack of metadata, the entire ecosystem of freebase.com web applications, application programming interfaces (APIs), and the applications that used the knowledge base's unique Metaweb Query Language (MQL) are also no longer available \cite{Bollacker2008}. Without such contextual information available today, it is important to consider that the Freebase data dumps represents a limited portion of the full knowledge base data and context.

\newpage{}

\bibliography{biblio}
\bibliographystyle{acm}

\newpage{}

\appendix

\section{Freebase Slices}

All of the Freebase domains have been clustered into one of three groups: Freebase Implementation Domains, OWL Domains, and Subject Matter Domains. The Freebase Implementation Domains include the triples that express the ontology and other helper domains that implemented Freebase on a technical level. OWL Domains include the triples based on the standard Web Ontology Language (OWL). Subject Matter Domains cover various domain areas on different subject matters in the world, from football to zoos.

\begin{longtable}{ | l | l | l | l | l | l | }
\caption{Freebase Domains} \\ \hline
No. & Name & Domain & Triples & Total \% &  Group \% \\ \hline
\multicolumn{6}{|l|}{\textit{Freebase Implementation Domains}} \\ \hline
1 & common & /common/* & 1,429,443,085 & 45.658\% & 58.507\% \\ \hline
2 & type & /type/* & 788,652,672 & 25.191\% & 32.280\% \\ \hline
3 & key & /key/* & 149,564,822 & 4.777\% & 6.122\% \\ \hline
4 & kg & /kg/* & 30,689,453 & 0.980\% & 1.256\% \\ \hline
5 & base & /base/* & 24,063,303 & 0.769\% & 0.985\% \\ \hline
6 & freebase & /freebase/* & 11,259,415 & 0.360\% & 0.461\% \\ \hline
7 & dataworld & /dataworld/* & 7,054,575 & 0.225\% & 0.289\% \\ \hline
8 & topic\_server & /topic\_server/* & 1,010,720 & 0.032\% & 0.041\% \\ \hline
9 & user & /user/* & 912,258 & 0.029\% & 0.037\% \\ \hline
10 & pipeline & /pipeline/* & 547,896 & 0.018\% & 0.022\% \\ \hline
11 & kp\_lw & /kp\_lw/* & 1,089 & 0.000\% & 0.000\% \\ \hline

\multicolumn{6}{|l|}{\textit{OWL Domains}} \\ \hline
1 & type & rdf-syntax-ns\#type & 266,321,867 & 8.507\% & 78.520\% \\ \hline
2 & label & rdf-schema\#label & 72,698,733 & 2.322\% & 21.434\% \\ \hline
3 & domain & rdf-schema\#domain & 71,338 & 0.002\% & 0.021\% \\ \hline
4 & range & rdf-schema\#range & 71,200 & 0.002\% & 0.021\% \\ \hline
5 & inverseOf & owl\#inverseOf & 12,108 & 0.000\% & 0.004\% \\ \hline
\multicolumn{6}{|l|}{\textit{Subject Matter Domains}} \\ \hline
1 & music & /music/* & 209,244,812 & 6.684\% & 60.062\% \\ \hline
2 & film & /film/* & 17,319,142 & 0.553\% & 4.971\% \\ \hline
3 & tv & /tv/* & 16,375,388 & 0.523\% & 4.700\% \\ \hline
4 & location & /location/* & 16,071,442 & 0.513\% & 4.613\% \\ \hline
5 & people & /people/* & 15,936,253 & 0.509\% & 4.574\% \\ \hline
6 & measurement\_unit & /measurement\_unit/* & 15,331,454 & 0.490\% & 4.401\% \\ \hline
7 & book & /book/* & 13,627,947 & 0.435\% & 3.912\% \\ \hline
8 & media\_common & /media\_common/* & 6,388,780 & 0.204\% & 1.834\% \\ \hline
9 & medicine & /medicine/* & 5,748,466 & 0.184\% & 1.650\% \\ \hline
10 & award & /award/* & 4,838,870 & 0.155\% & 1.389\% \\ \hline
11 & biology & /biology/* & 3,444,611 & 0.110\% & 0.989\% \\ \hline
12 & sports & /sports/* & 3,158,835 & 0.101\% & 0.907\% \\ \hline
13 & organization & /organization/* & 2,778,122 & 0.089\% & 0.797\% \\ \hline
14 & education & /education/* & 2,609,837 & 0.083\% & 0.749\% \\ \hline
15 & baseball & /baseball/* & 2,444,241 & 0.078\% & 0.702\% \\ \hline
16 & business & /business/* & 2,134,788 & 0.068\% & 0.613\% \\ \hline
17 & imdb & /imdb/* & 1,020,270 & 0.033\% & 0.293\% \\ \hline
18 & government & /government/* & 852,785 & 0.027\% & 0.245\% \\ \hline
19 & cvg & /cvg/* & 841,398 & 0.027\% & 0.242\% \\ \hline
20 & soccer & /soccer/* & 820,410 & 0.026\% & 0.235\% \\ \hline
21 & time & /time/* & 791,442 & 0.025\% & 0.227\% \\ \hline
22 & astronomy & /astronomy/* & 556,381 & 0.018\% & 0.160\% \\ \hline
23 & basketball & /basketball/* & 519,652 & 0.017\% & 0.149\% \\ \hline
24 & american\_football & /american\_football/* & 483,372 & 0.015\% & 0.139\% \\ \hline
25 & olympics & /olympics/* & 400,927 & 0.013\% & 0.115\% \\ \hline
26 & fictional\_universe & /fictional\_universe/* & 349,147 & 0.011\% & 0.100\% \\ \hline
27 & theater & /theater/* & 320,721 & 0.010\% & 0.092\% \\ \hline
28 & visual\_art & /visual\_art/* & 310,238 & 0.010\% & 0.089\% \\ \hline
29 & military & /military/* & 292,533 & 0.009\% & 0.084\% \\ \hline
30 & protected\_sites & /protected\_sites/* & 288,788 & 0.009\% & 0.083\% \\ \hline
31 & geography & /geography/* & 256,768 & 0.008\% & 0.074\% \\ \hline
32 & broadcast & /broadcast/* & 256,312 & 0.008\% & 0.074\% \\ \hline
33 & architecture & /architecture/* & 253,718 & 0.008\% & 0.073\% \\ \hline
34 & food & /food/* & 253,415 & 0.008\% & 0.073\% \\ \hline
35 & aviation & /aviation/* & 187,187 & 0.006\% & 0.054\% \\ \hline
36 & finance & /finance/* & 131,762 & 0.004\% & 0.038\% \\ \hline
37 & transportation & /transportation/* & 112,099 & 0.004\% & 0.032\% \\ \hline
38 & boats & /boats/* & 108,763 & 0.003\% & 0.031\% \\ \hline
39 & computer & /computer/* & 106,986 & 0.003\% & 0.031\% \\ \hline
40 & royalty & /royalty/* & 92,787 & 0.003\% & 0.027\% \\ \hline
41 & library & /library/* & 86,249 & 0.003\% & 0.025\% \\ \hline
42 & internet & /internet/* & 80,426 & 0.003\% & 0.023\% \\ \hline
43 & wine & /wine/* & 79,520 & 0.003\% & 0.023\% \\ \hline
44 & projects & /projects/* & 79,102 & 0.003\% & 0.023\% \\ \hline
45 & chemistry & /chemistry/* & 72,698 & 0.002\% & 0.021\% \\ \hline
46 & cricket & /cricket/* & 67,422 & 0.002\% & 0.019\% \\ \hline
47 & travel & /travel/* & 56,297 & 0.002\% & 0.016\% \\ \hline
48 & symbols & /symbols/* & 56,139 & 0.002\% & 0.016\% \\ \hline
49 & religion & /religion/* & 54,887 & 0.002\% & 0.016\% \\ \hline
50 & influence & /influence/* & 53,976 & 0.002\% & 0.015\% \\ \hline
51 & language & /language/* & 53,588 & 0.002\% & 0.015\% \\ \hline
52 & community & /community/* & 50,164 & 0.002\% & 0.014\% \\ \hline
53 & metropolitan\_transit & /metropolitan\_transit/* & 47,777 & 0.002\% & 0.014\% \\ \hline
54 & automotive & /automotive/* & 46,543 & 0.001\% & 0.013\% \\ \hline
55 & digicams & /digicams/* & 42,188 & 0.001\% & 0.012\% \\ \hline
56 & law & /law/* & 37,606 & 0.001\% & 0.011\% \\ \hline
57 & exhibitions & /exhibitions/* & 37,434 & 0.001\% & 0.011\% \\ \hline
58 & tennis & /tennis/* & 34,853 & 0.001\% & 0.010\% \\ \hline
59 & venture\_capital & /venture\_capital/* & 27,410 & 0.001\% & 0.008\% \\ \hline
60 & opera & /opera/* & 26,630 & 0.001\% & 0.008\% \\ \hline
61 & comic\_books & /comic\_books/* & 25,529 & 0.001\% & 0.007\% \\ \hline
62 & amusement\_parks & /amusement\_parks/* & 22,880 & 0.001\% & 0.007\% \\ \hline
63 & dining & /dining/* & 21,297 & 0.001\% & 0.006\% \\ \hline
64 & ice\_hockey & /ice\_hockey/* & 17,275 & 0.001\% & 0.005\% \\ \hline
65 & event & /event/* & 14,783 & 0.000\% & 0.004\% \\ \hline
66 & spaceflight & /spaceflight/* & 14,238 & 0.000\% & 0.004\% \\ \hline
67 & zoo & /zoo/* & 13,226 & 0.000\% & 0.004\% \\ \hline
68 & meteorology & /meteorology/* & 12,432 & 0.000\% & 0.004\% \\ \hline
69 & martial\_arts & /martial\_arts/* & 12,065 & 0.000\% & 0.003\% \\ \hline
70 & periodicals & /periodicals/* & 9,424 & 0.000\% & 0.003\% \\ \hline
71 & games & /games/* & 9,024 & 0.000\% & 0.003\% \\ \hline
72 & celebrities & /celebrities/* & 8,815 & 0.000\% & 0.003\% \\ \hline
73 & nytimes & /nytimes/* & 7,537 & 0.000\% & 0.002\% \\ \hline
74 & rail & /rail/* & 7,431 & 0.000\% & 0.002\% \\ \hline
75 & interests & /interests/* & 5,345 & 0.000\% & 0.002\% \\ \hline
76 & atom & /atom/* & 5,199 & 0.000\% & 0.001\% \\ \hline
77 & boxing & /boxing/* & 4,282 & 0.000\% & 0.001\% \\ \hline
78 & comic\_strips & /comic\_strips/* & 4,234 & 0.000\% & 0.001\% \\ \hline
79 & conferences & /conferences/* & 2,495 & 0.000\% & 0.001\% \\ \hline
80 & skiing & /skiing/* & 1,949 & 0.000\% & 0.001\% \\ \hline
81 & engineering & /engineering/* & 1,546 & 0.000\% & 0.000\% \\ \hline
82 & fashion & /fashion/* & 1,535 & 0.000\% & 0.000\% \\ \hline
83 & radio & /radio/* & 1,385 & 0.000\% & 0.000\% \\ \hline
84 & distilled\_spirits & /distilled\_spirits/* & 1,055 & 0.000\% & 0.000\% \\ \hline
85 & chess & /chess/* & 558 & 0.000\% & 0.000\% \\ \hline
86 & physics & /physics/* & 449 & 0.000\% & 0.000\% \\ \hline
87 & geology & /geology/* & 353 & 0.000\% & 0.000\% \\ \hline
88 & bicycles & /bicycles/* & 313 & 0.000\% & 0.000\% \\ \hline
89 & comedy & /comedy/* & 120 & 0.000\% & 0.000\% \\ \hline
\end{longtable}

\end{document}